\documentclass[prd,tightenlines,nofootinbib,showpacs,preprintnumbers,superscriptaddress, longbibliography, notitlepage]{revtex4-1}

\usepackage{amsfonts,amsmath,amssymb,amsthm,bbm,hyperref}
\usepackage{graphicx}
\usepackage{color}

\newcommand{\be}{\begin{equation}}
\newcommand{\ee}{\end{equation}}
\newcommand{\beq}{\begin{eqnarray}}
\newcommand{\eeq}{\end{eqnarray}}

\begin{document}

\title{Quantum cosmology in the light of quantum mechanics}
\author{Salvador J. Robles-P\'{e}rez}
\affiliation{Estaci\'{o}n Ecol\'{o}gica de Biocosmolog\'{\i}a, Pedro de Alvarado, 14, 06411 Medell\'{\i}n, Spain.}
\affiliation{Departamento de matem\'{a}ticas, IES Miguel Delibes, Miguel Hern\'{a}ndez 2, 28991 Torrej\'{o}n de la Calzada, Spain.}

\date{\today}

\begin{abstract}
There is a formal analogy between the evolution of the universe, when this is seen as a trajectory in the minisuperspace, and the worldline followed by a test particle in a curved spacetime. The analogy can be extended to the quantum realm, where the trajectories are transformed into wave functions that give us the probabilities of finding the universe or the particle in a given point of their respective spaces: the spacetime in the case of the particle and the minisuperspace in the case of the universe. The wave function of the spacetime and the matter fields, all together, can then be seen as a super-field that propagates in the minisuperspace and the so-called third quantisation procedure can be applied  in a parallel way as the second quantisation procedure is performed with a matter field that propagates in the spacetime. The super-field can thus be interpreted as made up of universes propagating, i.e. evolving, in the minisuperspace. The corresponding Fock space for the quantum state of the multiverse is then presented. The analogy can also be used in the opposite direction. The way in which the semiclassical state of the universe is obtained in quantum cosmology from the quantum state of the universe allows us to obtain, from the quantum state of a field that propagates in the spacetime, the geodesics of the underlying spacetime as well as their quantum uncertainties or dispersions. This might settle a new starting point for a different quantisation of the spacetime coordinates.
\end{abstract}

\pacs{98.80.Qc, 03.65.−w}
\maketitle

\tableofcontents

\section{Introduction}

In 1990 M. Gell-Mann and J. B. Hartle presented the sum-over-histories formulation of quantum cosmology in a paper entitled "Quantum mechanics in the light of quantum cosmology", in which the classical domains of familiar experience are derived from a decoherence process between the  alternative histories of the universe. In that paper  \cite{GellMann1990}, the authors conclude that  \emph{quantum mechanics is best and most fundamentally understood in the context of quantum cosmology}. This is so for many different reasons. First, the non-locality or, generally speaking, the non-separability of the quantum theory leads to the assumption that it cannot be applied only to \emph{a given system since it is not isolated but coupled to its natural environment, which is again coupled to another environment, and so forth} \cite{Kiefer1994}. The extrapolation of that idea inevitably implies that the quantum theory must be applied, from the most fundamental level, to the universe as a whole. If this is so, then, the quantum mechanics of particles and fields must be a derivable consequence of the application of the quantum theory to the whole universe.

For instance, there is no preferred time variable in the universe so, strictly speaking, it cannot undergo any time evolution. However, we know from experience that the spacetime and the things that are deployed in the spacetime evolve in time. Therefore, as Gell-Mann and Hartle show, time and time evolution, and particularly the Schr\"{o}dinger equation that provides us with the time evolution of quantum systems, must be emergent features of the quantum state of the universe. It is from that point of view from which the principles of quantum mechanics can be most fundamentally understood in the context of quantum cosmology, as the authors say. Furthermore, the universe jeopardises some of the fundamentals of the quantum theory. For instance, what does it mean concepts like uncertainty or non-locality in the context of a universe that is not deployed in the spacetime but it contains it? Thus, quantum cosmology forces us to acquire a deeper and a wider understanding of the quantum theory.

The idea behind quantum cosmology is that the conditions imposed on the state of the universe at the boundary\footnote{Time is created at the onset of the universe and thus the wave function of the universe cannot be a time-dependent function, so we cannot apply an initial condition on the state of the universe. However, the universe may  have a boundary where to impose the conditions that eventually would determine everything else in the whole history of the universe.}, together with the equations of quantum mechanics should be enough to assign probabilities to any plausible event that may happen in the universe. This is in principle the most that a non deterministic theory like the quantum mechanics can provide. With that purpose, and following a parallelism with the Feynman's formalism of path integrals, Gell-Mann and Hartle extend the seminal idea of Everett \cite{Everett1957} and develop their sum-over-histories theory \cite{GellMann1990, GellMann1993, Hartle1993}, in which a \emph{history} is defined as a time ordered sequence of projectors that represent all the possible outcomes that the infinite constituents of the universe may give at each moment of time\footnote{These are essentially the relative states of Everett's formulation of quantum mechanics \cite{Everett1957}. However, Everett did not provide an explanation of why some states and no others are selected from the whole set of possible states. In order to explain it Hartle needs to add, besides the boundary condition of the state of the universe and the equations of quantum mechanics, a new ingredient: the coarse-graining process that makes some states to emerge from the decoherence process. These are the selected states of the Everett's formulation.}. These fine-grained histories represent therefore all the possibilities in the universe and, hence, they contain all the information of the universe. However, these histories interfere among each other so in order to assign independent probabilities to the exclusive outcomes of the semiclassical experience\footnote{The outcomes of a classical experiment are exclusive, i.e. the cat is either dead or alive but not both.} one must take some coarse graining around the representatives values of the distinguished variables under study. In that process the fine detailed information is lost but it is because the loss of that ignored information that we can assign consistent probabilities to the alternative outcomes of a given experiment. It may seem then curious that the acceptance of a bit of ignorance is what allows us to obtain information from a physical system.

In the case of quantum cosmology, it turns out that (classical) time and the time evolution of matter fields, which constitute the main ingredients of the semiclassical domain of our everyday experience, are emergent features that decohere from the fine-detailed description of the  quantum state of the universe \cite{Hartle1993}. Thus, the sum-over-histories framework provides us with a consistent assignation of probabilities to the different outcomes of a given experiment and, in cosmology, it supplies us with an explanation for the appearance of the semiclassical domains of everyday experience\footnote{The existence of a semiclassical domain in the universe, and actually our own existence, can be seen as two possible outcomes of \emph{the cosmological experiment}. As Hartle says \cite{Hartle1993}, \emph{we live in the middle of this particular experiment}.}, where it is developed the quantum field theories of matter fields; so, at least from the conceptual point of view, everything seemed to be settled in quantum cosmology. The idea that was left is that little else could be done. In order to understand quantum mechanically the primordial singularity we need a complete quantum theory of gravity, and short after the origin the inflationary process seems to blur any posible imprint of the quantum regime of the universe. Besides, everything that follows could be explained by the quantum mechanics of particles and fields that, in the light of quantum cosmology, are emergent features of the quantum state of the whole universe. Thus, quantum cosmology got stuck in the 90's\footnote{Two important exceptions are the developments made in loop quantum cosmology \cite{Rovelli2001} and the computation of next order gravitational corrections to the Schr\"{o}dinger equation made in Refs. \cite{Kiefer1991, Kiefer2012, Brizuela2016a, Brizuela2016b}. }.

Almost thirty years afterwards, the title of this paper becomes a humble tribute to Gell-Mann and Hartle, and to many other authors that made possible the development of quantum cosmology \cite{Wheeler1957, DeWitt1967, Hawking1982, Hawking1983, Hawking1984, Hawking1987, Vilenkin1982, Vilenkin1984, Vilenkin1986, Vilenkin1988, Vilenkin1994, Vilenkin1995, Hartle1983, Hartle1990, Hartle1993, Halliwell1985, Halliwell1987, Halliwell1989, Halliwell1990, Kiefer1987, Kiefer1992, Kiefer1994, Kiefer2007, GellMann1990, Wiltshire2003}, and particularly to P. Gonz\'{a}lez-D\'{\i}az, who figuratively introduced me to all of them. However, it  also suggests the idea that it might be the time now, like in Plato's cavern allegory, of doing the way back to that proposed by Gell-Mann and Hartle. Perhaps, it may be now quantum cosmology the one that can be benefit of a deeper insight in the light of the well known principles of the quantum mechanics of particles and fields. With that aim in mind,  we shall use the analogy between the spacetime and the minisuperspace, as well as the analogy between their quantum mechanical descriptions, to shed some light in both directions. In one direction, the analogy between quantum cosmology and the quantum theory of a field that propagates in the spacetime  provides us with a useful framework where to develop a quantum theory of the whole multiverse. In the other direction, the way in which the semiclassical state of the universe is obtained in quantum cosmology from the quantum state of the universe will allow us to obtain, from the quantum state of a field that propagates in the spacetime, the classical trajectories followed by test particles as well as their quantum uncertainties. This might settle a new viewpoint for a different quantisation of the spacetime coordinates.

The paper is outlined as follows. In Sec. \ref{sec02} we give a brief justification of the use of the minisuperspace instead of the whole superspace. In Sec. \ref{sec03} we develop the classical analogy between the evolution of the universe in the minisuperspace and the trajectory of a test particle in a curved spacetime. In Sec. \ref{sec04} we consider the wave function of the spacetime and the matter fields, all together, as a super-field that propagates in the minisuperspace. Then, a similar quantisation formalism to that made in a quantum field theory is applied and the super-field is then interpreted made us of universes propagating in the minisuperspace. In Sec. \ref{sec05} we describe the semiclassical regime of the universe derived from the solutions of the Wheeler-DeWitt equation and, analogously, we also obtain the trajectories of test particles in the spacetime from the semiclassical expansion of the solutions of the Klein-Gordon equation. Finally, in Sec. \ref{sec06} we summarise and draw some conclusions.


\section{Brief justification of the minisuperspace}\label{sec02}

\begin{figure}
\centering
\includegraphics[width=14 cm]{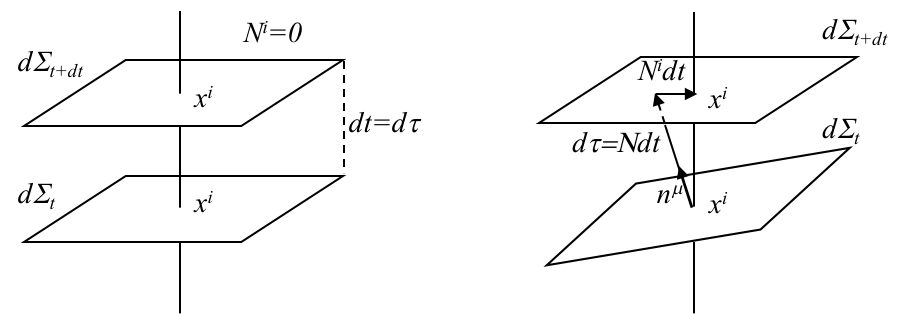}
\caption{Foliation of the spacetime into space and time. Left: in flat spacetime the lines of constant $x^i$ are orthogonal to the spatial hypersurfaces $\Sigma$ and the coordinate time $t$ coincides with the proper time, $\tau$. Right: in a curved spacetime, there is a shift, given by $N^i dt$, between the point that would have been reached if the particle would have followed the orthonormal vector of the hypersurface, $n^\mu$ at $x^i$ in $\Sigma_t$,  and the actual point of coordinates $x^i$ in $d\Sigma_{t+dt}$. The proper time $\tau$ is now 'lapsed' with respect to the coordinate time, $t$.}
\label{figure01}
\end{figure}

In the canonical approach of quantum cosmology, the quantum state of the universe is described by a wave function that depends on the variables of the spacetime and on the variables of the matter fields. It is the solution of the Wheeler-DeWitt equation \cite{DeWitt1967}, which is obtained by canonically quantising the Hamiltonian constraint associated to the classical Hilbert-Einstein action.  In order to see how complicated this can be let us briefly sketch the quantisation procedure. The first step consists of foliating the spacetime into space-like Cauchy hypersurfaces $\Sigma_t$, where $t$ denotes the global time function of the $3+1$ decomposition (see, Fig. \ref{figure01}). A line element of the spacetime can then be written as \cite{Wiltshire2003, Kiefer2007}
\be\label{STfol}
ds^2 = g_{\mu \nu} dx^\mu dx^\nu = \left(h_{ij} N^i N^j - N^2 \right) dt^2  + h_{ij} \left( N^i  dx^j + N^j dx^i \right) dt + h_{ij} dx^i dx^j ,
\ee
where $h_{ij}$ is the three-dimensional metric induced on each hypersurface $\Sigma_t$, given by \cite{Kiefer2007}
\be
h_{\mu\nu} = g_{\mu\nu} + n_\mu n_\nu ,
\ee
with the unit normal to $\Sigma_t$, $n_\mu$, satisfying, $n^\mu n_\mu = -1$, and $N$ and $N^i$ are called the lapse and the shift functions, respectively, which are the normal and tangential components of the vector field $t^\mu$, which satisfies $t^\mu \nabla_\mu t = 1$, with respect to the Cauchy hypersurface $\Sigma_t$ (see the details in, for instance, Refs \cite{Wiltshire2003, Kiefer2007}). In the Hamiltonian formulation of the Hilbert-Einstein action the variables of the phase space turn out to be then the metric components, $h_{ij}=h_{ij}(t,\vec x)$, and the conjugate momenta, which are functions of their time derivatives. The lapse and shift functions act as Lagrange multipliers. One eventually obtains a set of constraints that must be satisfied in the classical theory. The canonical procedure of quantisation consists of assuming the quantum version of the classical constraints. In particular, the Hamiltonian constraint, considering as well the variables of the matter fields $\varphi_n(t,\vec x)$, gives rise the well-known Wheeler-DeWitt equation \cite{DeWitt1967, Wiltshire2003, Kiefer2007},
\be\label{WDW00}
\left( - 16 \pi G \hbar^2 G_{ijkl} \frac{\delta^2}{\delta h_{ij}\delta h_{kl}} + \frac{\sqrt{h}}{16\pi G} \left(  -\, ^{(3)}R + 2 \Lambda + 16\pi G \hat T^{00} \right) \right) \phi(h_{ab}, \varphi) = 0 ,
\ee
where $G$, $\hbar$, and $\Lambda$ are the Newton, the Planck, and the cosmological constants, respectively, $h$ is the determinant of the metric $h_{ij}$, $^{(3)}R$ is the scalar curvature on the hypersurface $\Sigma_t$, $\hat T^{00}$ is the quantised version of the zero component of the energy momentum tensor of the matter field, which for instance, for a scalar field $\varphi$ reads
\be
\hat T^{00} = \frac{-1}{2 h} \frac{\delta^2}{\delta \varphi^2} + \frac{1}{2} h^{ij} \varphi_{,i} \varphi_{,j} + V(\varphi) ,
\ee
and $G_{ijkl}$ in (\ref{WDW00}) is called the supermetric \cite{DeWitt1967, Wiltshire2003, Kiefer2007},
\be
G_{ijkl} = \frac{1}{2\sqrt{h}} \left( h_{ik} h_{jl} + h_{il} h_{jk} - h_{ij}h_{kl} \right) . 
\ee
The wave function $\phi((h_{ab}, \varphi)$ in (\ref{WDW00}) is called the wave function of the universe \cite{Hartle1983}, and it is defined in the abstract space of all possible three-metrics defined in $\Sigma_t$ modulo diffeomorphisms, called the \emph{superspace}. Furthermore, the Wheeler-DeWitt equation (\ref{WDW00})  is not a single equation but in fact it is an equation at each point $x$ of the hypersurface $\Sigma_t$ \cite{Wiltshire2003}. It is then easy to understand that the exact solution of the Wheeler-DeWitt equation (\ref{WDW00}) is very difficult if not impossible to obtain for a general value of the metric and a general value of the matter fields. For practical purposes, one needs to assume some symmetries in the underlying spacetime to make it tractable. In that case, the number of variables of the superspace can be notably reduced and for that reason it is called the \emph{minisuperspace}\footnote{Note however that this space can still be infinite dimensional.}.

Furthermore, the observational data indicate that the most part of the history of the universe this is homogeneous and isotropic, at least as a first approximation. It seems therefore reasonable to consider the minisuperspace of the homogeneous and isotropic spacetimes instead of the full superspace. It is true that it would seem meaningless not to consider the full superspace to describe the quantum state of the universe since, at first, the most relevant regime for quantum cosmology seems to be the singular origin of the universe, where the quantum fluctuations of the spacetime make impossible to consider not only any symmetry of the spacetime but also the classical spacetime itself. However, this is not necessarily the case. First, to describe the initial singularity, if this exists, we would need a full quantum theory of gravity, which is not yet available. Second, there are homogeneous and isotropic models of the universe for which the scale factor does not contract further than a minimal value, $a_{\rm min}$. In that case, if $a_{\rm min}$ is of some orders higher than the Planck length the quantum fluctuations of the spacetime can clearly be subdominant. Furthermore, it could well happen that the quantum description of the universe would also be relevant at other scales rather than the Planck length, even in a macroscopic universe like ours. For all those reasons, we can consider  a homogeneous and isotropic spacetime as a first approximation and we can then analyse the small inhomogeneities of the spacetime and the matter fields as corrections to the homogeneous and isotropic background. This provides us with a relatively simple but still complete and useful model of the universe, at least for the major part of its evolution, even in quantum cosmology.

If one assumes isotropy, the metric of the three-dimensional hypersurfaces, $h_{ij}(t, \vec x)$ and the value of the matter fields, $\varphi_n(t, \vec x)$ can be expanded in spherical harmonics as \cite{Halliwell1985, Kiefer1987}
\beq\label{hdecomp}
h_{ij}(t,\vec x)  &=&  a^2 \Omega_{ij} + a^2  \sum_\textbf{n} 2 d_\textbf{n}(t) G^\textbf{n}_{ij}(\vec x) + \ldots  , \\ \label{fidecomp}
\varphi(t,\vec x)  &=&\frac{1}{\sqrt{2\pi}} \varphi(t) + \sum_\textbf{n} f_\textbf{n}(t) Q^\textbf{n}(\vec x) ,
\eeq
where $\Omega_{ij}$ are the metric components of a line element in the three-sphere, $Q^\textbf{n}(\textbf{x})$ are the scalar harmonics, and  $G^\textbf{n}_{ij}(\textbf{x})$ are the transverse traceless tensor harmonics, with $\textbf{n}\equiv(n,l,m)$ (see Ref. \cite{Halliwell1985} for the details). More terms appear in the expansion of the metric tensor \cite{Halliwell1985}. However, the dominant contribution is given by the tensor modes of the spacetime, $d_n$, and the scalar modes of the perturbed field, $f_n$, so let us focus on $d_n$ and $f_n$ as the representative of the inhomogeneous modes of the metric and matter fields, respectively\footnote{Eventually, these inhomogeneous modes can be interpreted as particles and gravitons propagating in the homogeneous and isotropic background spacetime.}.

If, as a first approximation, we only consider the homogeneous modes, the evolution of the universe is essentially described by two variables, the scale factor $a(t)$ and the homogeneous mode of the scalar field $\varphi(t)$. In that case, the Wheeler-DeWitt equation and the wave function of the universe depend only on these two variables, $(a, \varphi )$,  which turn out to be the coordinates of the corresponding minisuperspace. Although this minisuperspace may look a very simplified space it provides us with a very powerful context where to describe the evolution of a quite realistic model of the universe. Furthermore, it can easily be generalised without loosing effectiveness.  For instance, we could consider $n$ scalar fields\footnote{Spinorial and vector fields can be considered as well.}, $\varphi_1, \ldots, \varphi_n$, to represent the matter content of the universe. In that case, the resulting minisuperspace would be generated by the coordinates $(a, \vec \varphi)$, where $\vec \varphi=(\varphi_1, \ldots, \varphi_n)$. One could also consider anisotropies by choosing a minisuperspace with coordinates $(a_x, a_y, a_z, \vec \varphi)$, or we can consider isotropy but not homogeneity, as in (\ref{hdecomp}-\ref{fidecomp}), and then, the minisuperspace would be the infinite-dimensional space spanned by the variables $(a, \varphi, \vec f, \vec d, \ldots)$, where $\vec f = (f_1, f_2, \ldots)$, $\vec d = (d_1, d_2, \ldots)$, $\ldots$,  are the vectors formed with all the inhomogeneous modes of the expansions (\ref{hdecomp}-\ref{fidecomp}). It is easy to see then that the use of the minisuperspace is well justified to describe many models of  the universe, including the most realistic ones.

Furthermore, in all these cases the evolution of the universe can be seen as a parametrised trajectory in the corresponding minisuperspace, with parametric coordinates $(a(t), \varphi(t), \vec f(t), \vec d(t), \ldots)$, where the time variable $t$ is the parameter that parametrises the trajectory in the minisuperspace. In this paper, we shall assume an homogeneous and isotropic background as a first approximation and the inhomogeneities will be analysed as small perturbations propagating in the isotropic and homogeneous background. In that case, the evolution of the universe is basically given by the path in the $(a, \varphi)$ plane, and the inhomogeneities would only produce small vibrations in the other planes around the main trajectory in the minisuperspace. Until the advent of a satisfactory quantum theory of gravity, this seems to be the most we can consider rigorously in quantum cosmology. Even though, as we shall see in this paper, we can still obtain a lot of information and a deep insight within these models. For instance, they will allow us to uncover an accurate relationship that exists between the quantisation of the evolution of the universe in quantum cosmology and the well-known procedure of quantisation of  the trajectories of particles and matter fields in the spacetime.


\section{Classical analogy: the geometric minisuperspace}\label{sec03}

As we already pointed out, the evolution of the universe can be seen as a parametrised trajectory in the minisuperspace, with the time variable $t$ being the parameter that parametrised the trajectory. If we assume homogeneity and isotropy in the background spacetime, then, $N^i = 0$, $\forall i$ in (\ref{STfol}), so the metric  becomes 
\be\label{FRWmetric}
ds^2 = - N^2 dt^2 + a^2(t) d\Omega^2_3 ,
\ee 
where $a(t)$ is the scale factor, and $d\Omega^2_3$ is the line element on the three sphere. The lapse function $N$  parametrises here the ways in which the homogeneous and isotropic spacetime can be foliated into space and time, which are just time reparametrizations. If $N=1$ the time variable $t$ is called cosmic time and  if $N=a(t)$, $t$ is renamed with the Greek letter $\eta$, and it is then called conformal time because in terms of the time variable $\eta$ the metric becomes conformal to the metric of a closed static spacetime. For the matter fields, we consider the homogeneous mode of a scalar field, $\varphi(t)$, minimally coupled to gravity. Later on we shall  consider as well inhomogeneities as small perturbations of the homogeneous and isotropic background represented by (\ref{FRWmetric}). The Einstein-Hilbert action plus the action of the matter fields can then be written as \cite{Kiefer2007}
\be\label{S01}
S = S_g + S_m = \int dt N \left( \frac{1}{2} G_{AB} \frac{\dot{q}^A \dot q^B}{N^2}  - \mathcal V(q) \right) ,
\ee
where the variables of the minisuperspace, $q^A$, are the scale factor and the scalar field\footnote{\label{fn06} For convenience the scalar field has been rescaled according to, $\varphi \rightarrow \sqrt{2} \varphi$.}, i.e. $q \equiv \{a, \varphi\}$. The metric $G_{AB}$ of the minisuperspace, called the minisupermetric, is given in the present case by \cite{Kiefer2007}
\be\label{MSM01}
G_{AB} = {\rm diag}(-a, a^3) ,
\ee 
and the potential term, $\mathcal V(q)$ in (\ref{S01}), contains all the non kinetic terms of the action,
\be\label{V301}
\mathcal V(q) \equiv \mathcal V(a,\varphi) = \frac{1}{2} \left( - a + a^3 V(\varphi)  \right) .
\ee
The first term in (\ref{V301}) comes from the closed geometry of the three space, and $V(\varphi)$ is the potential of the scalar field. The case of a spacetime with a cosmological constant, $\Lambda$, is implicitly included if we consider a constant value of the potential of the scalar field, $V(\varphi)=\frac{\Lambda}{3}$.

\begin{figure}
\centering
\includegraphics[width=8 cm]{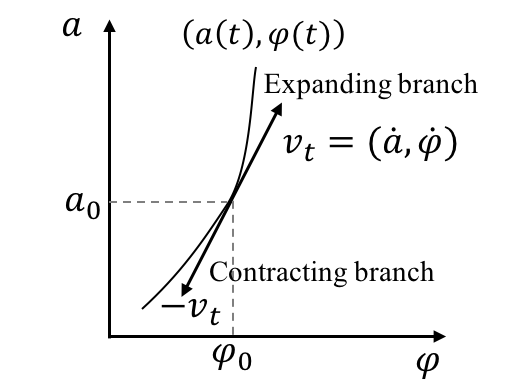}
\caption{The evolution of the universe can be seen as a parametrised trajectory in the minisuperspace. Trajectories in the minisuperspace with positive zero components of the tangent vector entail a growing value of the scale factor so they represent expanding universes. Analogously, those with negative zero component in the tangent vector describe contracting universes.}
\label{figure02}
\end{figure}

The action (\ref{S01}) and the minisupermetric (\ref{MSM01}) show that the minisuperspace is equipped with a geometrical structure formally similar to that of a curved spacetime. In the spacetime, the trajectory followed by a test particle is given by the path that extremizes the action \cite{Garay2018}
\be\label{ACTst}
S = \frac{m}{2} \int d\tau \, n \left( \frac{1}{n^2} g_{\mu\nu} \frac{dx^\mu}{d\tau}  \frac{dx^\nu}{d\tau} - m \right) , 
\ee
where $n$ is a function that makes the action (\ref{ACTst}) invariant under reparametrizations of the affine parameter $\tau$. The variation of the action (\ref{ACTst}) yields the well-known geodesic equation,
\be\label{GEOst}
\frac{d^2x^\mu}{d\tau^2} + \Gamma^\mu_{\alpha\beta} \frac{dx^\alpha}{d\tau} \frac{dx^\beta}{d\tau} = 0 ,
\ee
which in the Lagrangian formulation given by the action (\ref{ACTst}) turns out to be the Euler-Lagrange equation. Similarly, the classical evolution of the universe from the boundary states $(a_0, \varphi_0)$ and $(a_f,\varphi_f)$, can  be seen as the path that joins these two points of the minisuperspace and extremizes the action (\ref{S01}). The parametric coordinates $a(t)$ and $\varphi(t)$ of the curve that describes the evolution of the universe are then given by   
\be\label{EM01}
\ddot{q}^A + \Gamma^A_{BC} \dot{q}^B \dot{q}^C = - G^{AB} \frac{\partial \mathcal{V}}{\partial q^B} ,
\ee
where\footnote{Unless otherwise indicated we shall always consider cosmic time, for which $N=1$.}, $\dot{q}^A \equiv \frac{d q^A}{N dt}$, and $\Gamma^A_{BC}$ are the Christoffel symbols associated to the minisupermetric $G_{AB}$, defined as usual by
\be
\Gamma^A_{BC} = \frac{G^{AD}}{2} \left\{ \frac{\partial G_{BD}}{\partial q^C} + \frac{\partial G_{CD}}{\partial q^B} - \frac{\partial G_{BC}}{\partial q^D} \right\} .
\ee
In the case of the minisupermetric (\ref{MSM01}) the non vanishing components of $\Gamma^A_{BC}$ are
\be\label{CS01}
\Gamma^a_{aa} = \frac{1}{2a} \ , \ \Gamma^a_{\varphi\varphi} = \frac{3a}{2} \ , \ \Gamma^\varphi_{\varphi a} = \Gamma^\varphi_{a \varphi} = \frac{3}{2 a} .
\ee
Inserting (\ref{CS01}) in (\ref{EM01}) one obtains\footnote{Recall that the scalar field $\varphi$ has been rescaled according to $\varphi \rightarrow \sqrt{2}\varphi$, see f.n. \ref{fn06}.}
\be
\ddot{a} +\frac{\dot{a}^2}{2 a} + \frac{3a}{2} \dot \varphi^2 = -\frac{1}{2a} + 3 a V(\varphi) \ \ , \ \ 
\ddot \varphi + 3\frac{\dot a}{a} \dot \varphi = -\frac{\partial V(\varphi)}{\partial \varphi} ,
\ee
which are the classical field equations \cite{Linde1993, Kiefer2007}. The evolution of the universe can then be seen as a trajectory in the minisuperspace formed by the variables $a$ and $\varphi$ (see, Fig. \ref{figure02}). The time variable parametrises the \emph{worldline} of the universe and the solutions of the field equations, $a(t)$ and $\varphi(t)$, are the parametric coordinates of the universe along the worldline. Because the presence of the potential $\mathcal V$ in (\ref{EM01}), $t$ is not an affine parameter of the minisuperspace and the curved $(a(t), \varphi(t))$ is not an affinely parametrised geodesic. However, it is worth noticing that the action (\ref{S01}) is invariant under time reparametrisations. Therefore, we can make the following change in the time variable
\be
d\tilde t = \mathfrak m^{-2} \mathcal{V}(q)  d t ,
\ee
where $\mathfrak m$ is some constant. Together with the conformal transformation
\be
\tilde G_{AB} = \mathfrak m^{-2} \mathcal{V}(q) G_{AB} ,
\ee
the action (\ref{S01}) transforms as
\be\label{S02}
S = \int d\tilde t   N \left( \frac{1}{2  N^2} \tilde G_{AB} \frac{dq^A}{d\tilde t}  \frac{dq^B}{d\tilde t}  -  \mathfrak m^{2} \right)     .
\ee
The new time variable, with $ N=1$, turns out to be the affine parameter of the minisuperspace geometrically described by the metric tensor $\tilde G_{AB}$, with geodesics given by
\be\label{EM02}
\frac{d^2q^A}{d\tilde t^2} + \tilde \Gamma^A_{BC} \frac{dq^B}{d\tilde t} \frac{dq^C}{d\tilde t} = 0 .
\ee
Thus, the classical trajectory of the universe can equivalently be seen as a geodesic of the minisuperspace geometrically determined by the minisupermetric $\tilde G_{AB}$.

In the Lagrangian formulation of the trajectory of a test particle in the spacetime we can define the momenta conjugated to the spacetime variables as, $p_\mu = \frac{\delta \mathcal L}{\delta \frac{dx^\mu}{d\tau}}$. The invariance of the action (\ref{ACTst}) under reparametrisations of the affine parameter leads to the Hamiltonian constraint, $\frac{\delta H}{\delta n} = 0$, which turns out to be the momentum constraint of the particle
\be\label{HCst}
g^{\mu\nu} p_\mu p_\nu + m^2 = 0 .
\ee

A similar development can be done in the minisuperspace. The momenta conjugated to the variables of the minisuperspace are given by
\be
\tilde p_A \equiv \frac{\delta L}{\delta \frac{dq^A}{d\tilde t}} ,
\ee
and the Hamiltonian constraint associated to the action (\ref{S02}) turns out to be
\be\label{HC01}
\tilde G^{AB} \tilde p_A \tilde p_B + \mathfrak m^2 = 0 ,
\ee
or in terms of the metric $G_{AB}$ and the time variable $t$,
\be\label{HC02}
G^{AB}  p_A  p_B + \mathfrak m^2_{\rm ef}(q) = 0 ,
\ee
where for convenience we have written, $\mathfrak m^2_{\rm ef}(q) = 2 \mathcal V(q)$, with $\mathcal V(q)$ given by (\ref{V301}). It is worth noticing that the phase space does not change in the transformation $\{G_{AB}, t\} \rightarrow \{\tilde G_{AB}, \tilde t \}$, because
\be
\frac{\delta \tilde L}{\delta \left(\frac{dq^A}{d\tilde t}\right)} \equiv \tilde p_A   = \tilde G_{AB} \frac{d q^B}{d\tilde t} = G_{AB} \frac{dq^B}{d t} = p_A \equiv  \frac{\delta  L}{\delta \left(\frac{dq^A}{d t}\right)}    ,
\ee
where, $p_A = \{p_a, p_\varphi\}$ and $q^A \equiv \{a, \varphi\}$.

There is then a clear analogy between the evolution of the universe, when this is seen as a path in the minisuperspace, and the trajectory of a test particle that moves in a curved spacetime. It allows us not only to see the evolution of the universe as a trajectory in the minisuperspace but also to attain a better understanding of  the quantisation of both the evolution of the universe and the trajectory of a particle in the spacetime. Within the former, we shall see that this analogy allows us to consider the wave function of the universe as another field, say a \emph{super-field}, that propagates in the minisuperspace, and whose quantisation can thus follow a similar procedure to that employed in the quantisation of a matter field that propagates in the spacetime. Therefore, following the customary interpretation made in a quantum field theory, this  new field can be interpreted in terms of \emph{test particles} propagating in the minisuperspace, i.e. universes evolving according to their worldline coordinates. From this point of view, the natural scenario in quantum cosmology is then a many-universe system, or multiverse. In the opposite direction in the relationship between the minisuperspace and the spacetime, the way in which the semiclassical description of the universe is obtained, i.e. the way in which a classical trajectory in the minisuperspace is recovered from the quantum state of the wave function of the universe, will allow us to recover from the quantum state of the field $\varphi$ the geodesics of the spacetime where it propagates, i.e. the trajectories followed by the particles of the field, as well as the uncertainties or deviations from their classical trajectories, given by the corresponding Schr\"{o}dinger equation.


\section{Quantum picture}\label{sec04}

\subsection{Quantum field theory in the spacetime}

The formal analogy between the minisuperspace and a curved spacetime can be extended to the quantum picture too. Let us first notice that in the quantum mechanics of fields and particles, the momentum constraint (\ref{HCst}) can be quantised by transforming the momenta conjugated to the spacetime variables into operators, $p_\mu \rightarrow -i \hbar \frac{\partial}{\partial x^\mu}$. With an appropriate choice of factor ordering\footnote{The one that makes the Klein-Gordon equation (\ref{KG01}) invariant under rotations in the spacetime.}, it gives the so-called Klein-Gordon equation
\be\label{KG01}
\left( -\hbar^2 \Box_x + m^2 \right) \varphi(t, \vec x) = 0 ,
\ee
where,
\be
\Box_x \varphi =  g^{\mu \nu} \nabla_\mu \nabla_\nu \varphi = \frac{1}{\sqrt{-g}} \frac{\partial}{\partial x^\mu} \left( \sqrt{-g} g^{\mu\nu} \frac{\partial  \varphi }{\partial x^\nu}\right) ,
\ee
with, $g={\rm det}(g_{\mu\nu})$. Note however the presence in (\ref{KG01}) of the Planck constant, $\hbar$, which does not appear when the Klein-Gordon equation is derived from the action of a classical field\footnote{This way of obtaining the Klein-Gordon equation from the Hamiltonian constraint of a test particle that propagates in the spacetime is well-know since a long time. It can be seen, for instance, in Ref. \cite{Chernikov1968}. However it is not customary used in quantum field theory. The implications of the appearance of the Planck constant in the Klein-Gordon equation can be quite relevant. It will be shown in Sec. \ref{sec05} and they are extensively analysed  in Ref. \cite{Garay2018}.}. The field $\varphi(t,\vec x)$ in (\ref{KG01}) is then interpreted as a field that propagates in the spacetime, which is the configuration space of the Klein-Gordon equation. If we were just considering one single particle, then, the Klein-Gordon equation (\ref{KG01}) would be enough to provide us with the classical and the quantum description of the particle. Let us notice that the classical trajectory of the particle, as well as the uncertainties in the position given by the Schr\"{o}dinger equation, are already contained in the Klein-Gordon equation (\ref{KG01})  (see, Sect. \ref{sec05} and Ref. \cite{Garay2018}).

However, the most powerful feature of a quantum field theory is that it allows us to describe the quantum state of a many-particle system, and there, in the many particle scenario, new quantum effects can appear that cannot be present in the context of one single particle, like entanglement and other quantum correlations. Therefore, let us consider the so-called \emph{second quantisation} procedure of the scalar field $\varphi$, which  follows as it is well known (see, for instance, Refs. \cite{Birrell1982, Mukhanov2007}) by expanding the field $\varphi(x)$ in normal modes $u_k(x)$,
\be\label{vphi01}
\varphi(x) = \sum_k a_k u_k(x) + a^\dag_k u_k^*(x) ,
\ee
where the modes $u_k(x)$ are orthonormal in the product \cite{Birrell1982}
\be
(\varphi_1,\varphi_2) = - i \int_\Sigma \varphi_1(x)  \overset{\leftrightarrow}{\partial}_\mu \varphi^*_2(x) \sqrt{h} d\Sigma^\mu ,
\ee
where $d\Sigma^\mu = n^\mu d\Sigma$, with $n^\mu$ a future-directed unit vector orthogonal to the three-dimensional hypersurface $\Sigma$, $d\Sigma$ is the volume element in $\Sigma$, and $h$ is the determinant of the metric induced in $\Sigma$, i.e. $h={\rm det}(h_{ij})$. In that case, the modes $u_i(x)$ satisfy the customary relations
\be
(u_k, u_l) = \delta_{kl} , \,\, (u_k^*, u_l^*) = - \delta_{kl} , \,\, (u_k, u_l^*) = 0 .
\ee
The quantisation of the field (\ref{vphi01}) is then implemented by adopting the commutation relations
\be
[a_k, a_l^\dag] = \delta_{kl}, \,\, [a_k, a_l ] = [a_k^*, a_l^*] = 0 .
\ee
Then, one defines a vacuum state, $|0\rangle = \prod_k |0_k\rangle$, where $|0_k\rangle$ is the state annihilated by the $a_k$ operator, i.e. $a_k |0_k \rangle = 0$. The vacuum state $|0_k\rangle $ describes, in the representation defined by $a_k$ and $a_k^\dag$, the no-particle state for the mode $k$ of the field. We can then define the excited state,
\be
|m_{k_1}, n_{k_2}, \ldots \rangle = \frac{1}{\sqrt{m! n! \ldots}} \left( \left( a_{k_1}^\dag \right)^m \left( a_{k_2}^\dag \right)^n \ldots \right) |0\rangle ,
\ee
as the many-particle state representing $m$ particles in the mode $k_1$, $n$ particles in the mode $k_2$, etc. The definition of the Fock space is a very important step of the quantisation procedure and it allows us to write the general quantum state of the field as
\be\label{QSfield}
|\varphi\rangle = \sum_{m,n,\ldots} C_{m,n,\ldots} | m_{k_1} n_{k_2} \ldots \rangle ,
\ee 
where $|C_{m,n,\ldots}|^2$ is the probability to find $m$ particles in the mode $k_1$, $n$ particles in the mode $k_2$, etc. Thus, the field can be interpreted as made up of particles propagating in the spacetime with different values of their momenta. The quantum state of the field (\ref{QSfield}) contains all the power of the quantum field theory. For instance, it allows us to consider an entangled state like
\be\label{ES01}
| \varphi \rangle = \sum_n C_n | n_{\vec k}, n_{-\vec k} \rangle  = C_0 | 0_{\vec k}, 0_{-\vec k} \rangle + C_1 | 1_{\vec k}, 1_{-\vec k} \rangle + \ldots ,
\ee
which represents the linear combination of perfectly correlated pairs of particles moving in opposite directions (with opposite values of their spatial momenta, $\vec k$ and $-\vec k$). An entangled state like (\ref{ES01}) revolutionised the quantum mechanics, it showed that the \emph{distinguishing feature} of quantum mechanics  is the non-locality, or better said the non-separability, of the quantum states \cite{Schrodinger1936b, Bell1987}. It also entailed the appearance of new crucial developments in the physics of nowadays like, for instance, quantum information theory and quantum computation, among others. In the case of the universe, it seems now quite bizarre to think of a similar step  in quantum cosmology. However, if the expected effects  \cite{RP2018a, RP2018b} would be confirmed by astronomical observation, it would certainly revolutionise the  picture of our universe in a similar way.

\subsection{Quantum field theory in the minisuperspace}

A similar procedure of canonical quantisation can be followed in the minisuperspace by establishing the correspondence principle between the quantum and the classical variables of the phase space when they are applied upon the wave function, $\phi = \phi(a,\varphi)$. In the configuration space,
\be
a \rightarrow \hat a = a \phi  , \varphi \rightarrow \hat{\varphi} = \varphi \phi  \ , \ p_{a} \rightarrow \hat{p}_{a} \equiv - i \hbar \frac{\partial \phi }{\partial a}   , p_\varphi \rightarrow \hat{p}_\varphi \equiv - i \hbar \frac{\partial \phi}{\partial \varphi} .
\ee
Then, with an appropriate choice of factor ordering, the Hamiltonian constraint (\ref{HC01}) transforms into the Wheeler-DeWitt equation
\be\label{QHC01}
 \left( -\hbar^2 \tilde \Box_q  + \mathfrak m^2 \right) \phi = 0 ,
\ee
with, $\tilde\Box_q \equiv \tilde \nabla^2_{LB}$, where the \emph{Laplace-Beltrami operator} $\nabla_{LB}$ is the covariant generalisation of the Laplace operator \cite{Kiefer2007}, given by
\be\label{LB01}
\tilde \Box_q \equiv \tilde\nabla^2_{LB} = \frac{1}{\sqrt{-\tilde G}} \partial_A \left( \sqrt{-\tilde G} \tilde G^{AB} \partial_B \right) ,
\ee
or in terms of the variables without tilde the classical Hamiltonian constraint (\ref{HC02}) becomes
\be\label{QHC02}
 \left(  -\hbar^2 \Box_q  +  \mathfrak m^2_{\rm ef}(q) \right) \phi = 0 ,
\ee
where $ \Box_q$ is the Laplace-Beltrami operator (\ref{LB01}) with the metric $G_{AB}$ instead of  $\tilde G_{AB}$.

The customary approach of quantum cosmology consists of considering the solutions, exact or approximated, of the Wheeler-DeWitt equation (\ref{QHC02}) and to analyse the quantum state of the universe from the perspective of the wave function so obtained. This is what we can call the \emph{quantum mechanics of the universe} \cite{Hartle1990, Hartle1993}. This is the only thing we need if we are just considering the physics of one single universe, which has been the cosmological paradigm so far. As it is well-known (see, for instance, Refs. \cite{Hartle1990, Halliwell1990}), the wave function $\phi$ contains, at the classical level, the trajectory of the universe in the minisuperspace, i.e. the classical evolution of its homogeneous and isotropic background, and at first order in $\hbar$ it contains the Schr\"{o}dinger equation for the matter fields that propagate in the background spacetime. Thus, it contains in principle all the physics within a single universe.

However, as we have seen in the case of a field that propagates in the spacetime, it is the description of the field in a quantum field theory what extracts all the power of the quantum theory. We are then impeled  to follow a similar approach and exploit the remarkable parallelism between the geometric structure of the minisuperspace and the geometrical properties of a curved spacetime to interpret the wave function $\phi(a,\varphi)$ as a field that propagates in the minisuperspace. We shall then formally apply a procedure of quantisation that parallels that of a second quantisation, which is sometimes called \emph{third-quantisation} \cite{Caderni1984, McGuigan1988, Rubakov1988, Strominger1990, RP2010}  to be distinguished from the customary one. Then, let us go on with the parallelism and quantise the super-field $\phi(a,\varphi)$  by expanding it in terms of normal modes
\be\label{MEx01}
 \phi(q) = \sum_i \left(  b_i u_i(q) +  b_i^\dag u_i^*(q) \right) , 
\ee
where the index $i$ schematically represents the set of quantities necessary to label the modes, the sum must be understood as an integral for the continuous labels, and the functions $u_i(q)$ and $u_i^*(q)$ form now a complete set of mode solutions of the Wheeler-DeWitt equation (\ref{QHC01}), which are orthonormal under the product
\be\label{SP01}
(u_1(q) , u_2(q) ) = - i \int_\Sigma u_1(q) \overset{\leftrightarrow}{\partial}_\mu u_2^*(q) \sqrt{g_\Sigma} \ d\Sigma^\mu ,
\ee
where, in analogy to a curved spacetime \cite{Birrell1982}, $d\Sigma^\mu = n^\mu d\Sigma$, with $n^\mu$ a future directed unit vector\footnote{By a future directed vector in the minisuperspace we mean a vector positively oriented with respect to the scale factor component, which is the time-like variable of the minisuperspace.} orthogonal to the spacelike hypersurface $\Sigma$ in the minisuperspace, with induced metric given by $g_\Sigma$ and volume element  $d\Sigma$. Let us notice that the modes $u_i(q)$ in (\ref{MEx01})  depend now on the variables of the minisuperspace, $q^A=\{a, \varphi \}$, instead of on the coordinates of the spacetime. In the minisuperspace with minisupermetric $G_{AB}$, a natural choice is the 1-dimensional subspace generated at constant $a$ by the variable $\varphi$ ($d\Sigma = d\varphi$), then, $g_\Sigma = a^3$ and $n^\mu=(a^{-\frac{1}{2}}, 0)$, so the scalar product (\ref{SP01}) becomes \cite{RP2011b}
\be
(u_1 , u_2 ) = - i \int_{-\infty}^{+\infty} d\varphi \, a \ \left( u_1(a,\varphi) \overset{\leftrightarrow}{\partial}_a u_2^*(a,\varphi)  \right) .
\ee
The quantisation of the theory is then implemented by adopting the customary commutation relations between the mode operators $b_i$ and $b_i^\dag$, i.e. 
\be
[b_i, b_j^\dag] = \delta_{ij} \ , \ [b_i, b_j] = [b_i^\dag , b_j^\dag ] = 0 .
\ee
This is what we can call \emph{second quantisation} of the spacetime and the matter fields, all together\footnote{We do not call it second quantisation of the universe because in this formalism there is not only a universe but a set of many universes described analogously to the many-particle representation of a quantum field theory.}. The operators $b_i^\dag$ and $b_i$ are now the creation and the annihilation operators of universes, whose physical properties are described by the  solutions, $u_i(q)$, of the Wheeler-DeWitt equation (see Sec. \ref{sec05}). 

Then, similarly to a quantum field theory in the spacetime, we have to define a ground state, $|0\rangle = \prod_i |0_i\rangle$, where $| 0_i\rangle$ is the state annihilated by the operator $b_i$, i.e. $b_i |0_i\rangle = 0$. It describes, in the representation defined by $b_i$ and $b_i^\dag$, the no-universe state for the value $i$ of the mode. It means that the ground state $|0\rangle$ represents the no-universe at all state, which is sometimes called the \emph{nothing} state \cite{Strominger1990}. The excited state, i.e. the state representing different number of universes with values $i_1, i_2, \ldots$, is then given by
\be\label{FS01}
| m_{i_1}, n_{i_2}, \ldots \rangle = \frac{1}{\sqrt{m! n! \ldots}} \left[ \left( b^\dag_{i_1} \right)^m \left( b^\dag_{i_2} \right)^n \ldots \right]  |{}_b 0 \rangle ,
\ee
which represents $m$ universes in the mode $i_1$, $n$ universes in the mode $i_2$, etc. Let us notice that in the case of a field that propagates in a homogeneous and isotropic spacetime the value of the mode $\vec k$ represents the value of the spatial momentum of the particle \cite{Birrell1982, Mukhanov2007}. In the homogeneous and isotropic minisuperspace described by $G_{AB}$, it is the eigenvalue of the momentum conjugated to the scalar field $\varphi$, which formally plays the role of a spatial like variable in the minisuperspace. Therefore, the values $i_1, i_2, \ldots$, in (\ref{FS01}) label the different initial values of the time derivatives of the scalar field in the universes. Thus, the state (\ref{FS01}) represents $m$ universes with a scalar field with $\dot \varphi \sim i_1$, $n$ universes with a scalar field with $\dot \varphi \sim i_2$, etc. They represent different energies of the matter fields and, therefore, different number of particles in the universes. The general quantum state of the field $\phi$, which represents the quantum state of the spacetime and the matter fields, all together, is then given by 
\be\label{QSM01}
| \phi \rangle = \sum_{m,n,\ldots} C^{(b)}_{mn\ldots} | {}_b m_{i_1} n_{i_2} \ldots \rangle ,
\ee
which represents therefore the \emph{quantum state of the multiverse} \cite{RP2010}.

In the quantisation of a field that propagates in a curved spacetime there is an ambiguity in the choice of mode operators of the quantum scalar field. The different representations are eventually related by a Bogolyubov transformation so at the end of the day the vacuum state of one representation turns out to be full of particles\footnote{In the quantisation of a complex scalar field it would be full of particle-antiparticle pairs.} of another representation \cite{Mukhanov2007}. The ambiguity is solved by imposing the appropriate boundary conditions that give rise to the invariant representation, in which the vacuum state represents the no particle state along the entire history of the field \cite{RP2017d}. In the minisuperspace $b_i^\dag$ and $b_i$ in (\ref{MEx01}) would be the creation and the annihilation operators, respectively, of the corresponding invariant representation \cite{RP2017d}. Thus, the ground state of the invariant representation, $|0\rangle$, would represent the \emph{nothing state} at any point of the minisuperspace. It seems therefore to be the appropriate representation to describe the universes of the multiverse. However, it could well happen that the state of the super-field $\phi$ at the boundary $\Sigma(a_0)$, where $a_0$ is the value of the scale factor at which the universes are created from the gravitational vacuum,  would be  given by the ground state of the diagonal representation of the Hamiltonian at $a_0$, given by $\bar{b}_i^\dag$ and $\bar b_i$.  In terms of the invariant representation, the super-field $\phi$ would be then represented by an infinite number of universes, because \cite{Mukhanov2007}
\be\label{VS01}
|\bar 0 \rangle =\prod_i \frac{1}{|\alpha_i|^\frac{1}{2}}  \left( \sum_n \left( \frac{\beta_i}{2 \alpha_i} \right)^n | n_i, n_{-i} \rangle   \right) ,
\ee
where $\alpha_i$ and $\beta_i$ are the Bogolyubov coefficients that relate both representations, i.e.
\be
b_i = \alpha_i^* \bar b_i + \beta_i \bar b^\dag \,\,\, , \,\,\, b_i^\dag = \alpha_ i \bar b_i^\dag + \beta_i^* \bar b_i . 
\ee
It is worth noticing that because the isotropy of the underlying minisuperspace, the universes would be created in perfectly correlated states, $| n_i, n_{-i} \rangle $, with opposite values of their momenta, $i$ and $-i$. The creation of universes in pairs with opposite values of the momenta conjugated to the minisuperspace variables would conserved the value of the total momentum and it is besides a consequence of the quantum creation of universes in (\ref{VS01}). As we shall see in Sec. \ref{sec05} it will have important consequences because the time variables of the two universes of a given pair are reversely related \cite{RP2018c}. Therefore, particles propagating in the observer's universe would be clearly identified with matter and particles moving in the time reversely universe can naturally be identified with antiparticles. It might explain, therefore, the primordial matter-antimatter asymmetry observed in the context of a single universe \cite{RP2017e}.

\begin{figure}
\centering
\includegraphics[width=16 cm]{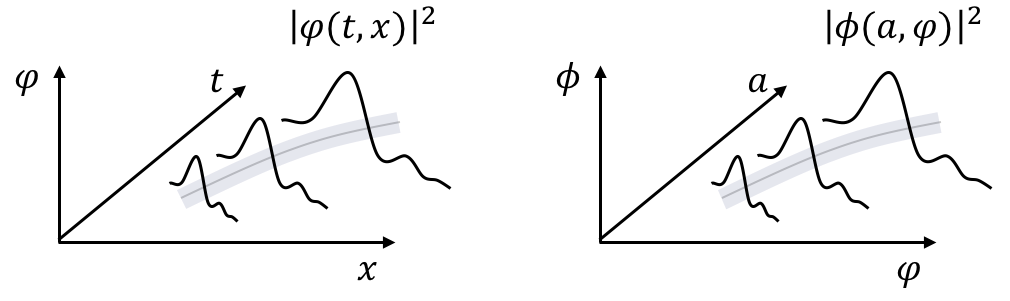}
\caption{Left: in a quantum field theory the field is described in terms of particles that follow with the highest probability the classical trajectories given by the geodesics with however some uncertainties in their positions. Right: the wave function that describes the quantum state of the spacetime and the matter fields, all together, can be seen as a another field, say a super-field, that propagates in the minisuperspace. The universes can then be seen as 'test' particles following classical trajectories in the minisuperspace with quantum uncertainties given by the Schr\"{o}dinger equation of their matter fields.}
\label{figure03}
\end{figure}

\section{Particles and universes propagating in their spaces}\label{sec05}

\subsection{Semiclassical universes: classical spacetime and quantum matter fields}

In quantum mechanics, the trajectories are transformed into wave packets. Instead of definite positions and definite trajectories, what we have in quantum mechanics is a wave function that gives us the probability of finding a particle in a particular point of the spacetime (see, Fig. \ref{figure03}). In the semiclassical regime this probability is highly peaked around the classical trajectory and we recover the picture of a classical particle propagating along the particle worldline.

Similarly, we can see quantum cosmology as the quantisation of the classical trajectory of the universe in the minisuperspace. In that case, the wave function $\phi(a,\varphi)$ can be interpreted as a field made up of universes which, in the classical limit, follow definite trajectories in the minisuperspace, i.e.  their spacetime backgrounds follow in that limit the classical evolution determined by the field equations. At first order in $\hbar$, however, there is some uncertainty in the matter field coordinates given by the Schr\"{o}dinger  equation of the matter fields.

\begin{figure}
\centering
\includegraphics[width=16 cm]{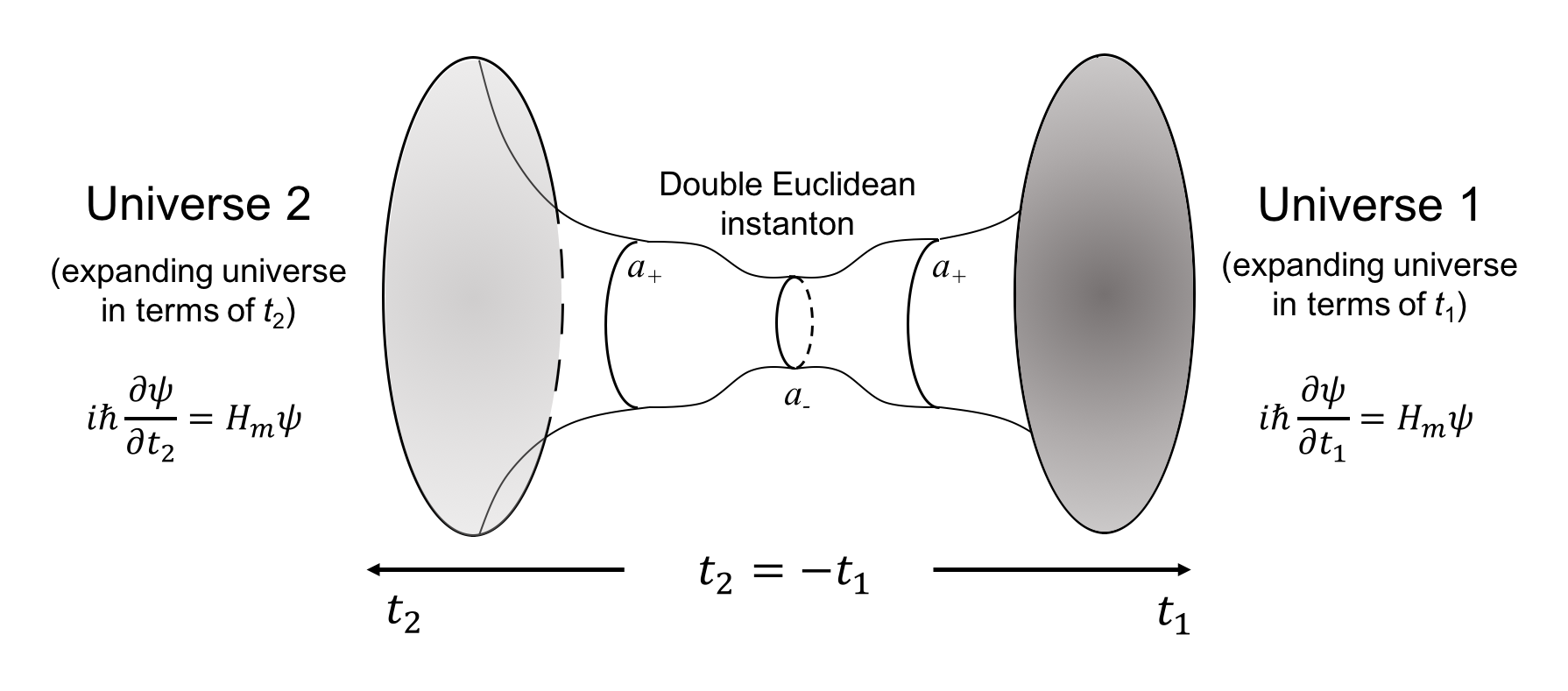}
\caption{The creation of universes in entangled pairs \cite{RP2018a}. In order to obtain the correct value of the Schr\"{o}dinger equation in the two universes, their physical time variables must be reversely related. In that case, particles moving in the symmetric universe look as they were moving backward in time so they are naturally identified with the antiparticles that are left in the observer's universe. The primordial matter-antimatter asymmetry observed in the context of a single universe would thus be restored in the  multiverse. Particles and antiparticles do not collapse at the onset because the Euclidean gap that exists between the two newborn universes \cite{RP2018a, RP2017e}.}
\label{figure04}
\end{figure}

In order to show it, let us consider the WKB solutions of the Wheeler-DeWitt equation (\ref{QHC02}), which can be written as
\be\label{WKB01}
\phi(q) = \sum_n C_n(q) e^{+\frac{i}{\hbar} S_n(q)} + C_n(q) e^{-\frac{i}{\hbar} S_n(q)} ,
\ee
where $C_n(q)$ and $S_n(q)$ are a slow-varying and a rapid varying functions, respectively, of the minisuperspace variables, and the sum extends to all possible classical configurations \cite{Hartle1990}. Because the real character of the Wheeler-DeWitt equation, which in turn is rooted on the time reversal symmetry of the Hamiltonian constraint (\ref{HC02}), the semiclassical solutions come in conjugate pairs like in (\ref{WKB01}). These two solutions represent classical universes is the following sense. If we insert them into the Wheeler-DeWitt equation (\ref{QHC02}) and expand it in power of $\hbar$, then, at zero order in $\hbar$ it is obtained the following Hamilton-Jacobi equation
\be\label{HJ01}
G^{AB} \frac{\partial S}{\partial q^A} \frac{\partial S}{\partial q^B} + \mathfrak m^2_{\rm ef}(q) = 0 .
\ee
It can be shown \cite{Hartle1990, Hartle1993} that this equation turns out to be the Hamiltonian constraint (\ref{HC02}) if we assume a time parametrisation of the paths in the minisuperspace given by
\be\label{WKBtime}
\frac{\partial }{\partial t} = \pm G^{AB} \frac{\partial S}{\partial q^A}   \frac{\partial }{\partial q^B}  .
\ee
In that case, 
\be\label{WKBmom01}
\dot q^A = \pm G^{AB} \frac{\partial S}{\partial q^B} , \,\,\,\,\,\, {\rm and} \,\,\,\,\,\,  \frac{\partial S}{\partial q^A} = \pm G_{AB} \dot{q}^B = p_A ,
\ee 
so that the Hamilton-Jacobi equation (\ref{HJ01}) becomes the Hamiltonian constraint (\ref{HC02}). Furthermore, from (\ref{WKBmom01}) and (\ref{HJ01}) one can derive the equation of the geodesic of the minisuperspace (\ref{EM01}). Therefore, at the classical level, i.e. in the limit $\hbar \rightarrow 0$, one recovers from the semiclassical solutions (\ref{WKB01}) the classical trajectory of the universe in the minisuperspace, i.e. one recovers the classical description of the background spacetime of the universe. In that sense, these solutions describe the classical spacetime of the universes they represent. It is worth noticing the freedom that we have to choose the sign of the time variable in (\ref{WKBtime}), $+t$ or $-t$. The Hamiltonian constraint (\ref{HC02}) is invariant under a reversal change in the time variable because the quadratic terms in the momenta. However, the value of these momenta in (\ref{WKBmom01}) is not invariant under the reversal change of the time variable. It means that we have two possible values of the momenta, $+p_A$ and $-p_A$, which are associated to the conjugated solutions of the Wheeler-DeWitt equation (\ref{QHC02}). It means that, as it happens in particle physics, the universes are created in pairs with opposite values of their  momenta so that the total momentum is conserved (see, Fig. \ref{figure04}). In the time parametrisation of the minisuperspace, the two reversely related time variables, $t$ and $- t$, represent the two possible directions in which the worldlines can be run in the minisuperspace, with positive and negative tangent vectors, $\pm v_t$ (see, Fig. \ref{figure02}). It means that one of the universes is moving forward and the other is moving backward in terms of the variables of the minisuperspace. One of these variables is the scale factor so, in particular, one of the universes is increasing the value of the scale factor, so it is an expanding universe, and the other is reducing the value of the scale factor, so it is a contracting universe.

At zero order in $\hbar$ in the expansion in powers of $\hbar$ of the Wheeler-DeWitt equation with the semiclassical states (\ref{WKB01}) it is obtained the classical background. At first order in $\hbar$ it is obtained the Schr\"{o}dinger equation of the matter fields that propagate in the background spacetime. Then, one recovers from the semiclassical states (\ref{WKB01}) the semiclassical picture of quantum matter fields propagating in a classical spacetime. For the shake of concreteness, let us consider the minisuperspace of homogeneous and isotropic spacetimes considered in Secs. \ref{sec03} and \ref{sec04}, with small inhomogeneities propagating therein. In these are small, the Hamiltonian of the background and the Hamiltonian of the inhomogeneities are decoupled, so the total Hamiltonian can be written \cite{Halliwell1985, Kiefer1987}
\be\label{HC03}
 (\hat{H}_{bg} + \hat{H}_m) \phi = 0 ,
\ee
where the Hamiltonian of the background spacetime, $H_{bg}$, is given by
\be\label{H0}
\hat{H}_{bg} =  \frac{1}{2 a} \left( \frac{\partial^2}{\partial a^2} + \frac{1}{a} \frac{\partial}{\partial a} - \frac{1}{a^2} \frac{\partial^2}{\partial \phi^2} + a^4 V(\varphi) - a^2  \right)  ,
\ee
and $H_m$ is the Hamiltonian of the inhomogeneous modes of the matter fields. In that case, the wave function $\phi$ depends not only on the variables of the background but also on the inhomogeneous degrees of freedom, i.e. $ \phi = \phi(a, \varphi; \vec x_\textbf{n})$, where $\vec x_\textbf{n}$ can denote either the tensor modes of the perturbed spacetime, $d_\textbf{n}$, or the scalar modes of the perturbed field, $f_\textbf{n}$ (see, (\ref{hdecomp}-\ref{fidecomp})). The semiclassical wave function (\ref{WKB01}) can now be written as \cite{Hartle1990, Kiefer1992}
\be\label{SCWF01}
\phi = \sum \phi_+ + \phi_- = \sum C e^{ \frac{i}{\hbar} S_0} \psi + C e^{ -\frac{i}{\hbar} S_0} \psi^*,
\ee
where $C$ and $S$ depend only on the variables of the background, $a$ and $\varphi$, and  $\psi=\psi(a, \varphi; \vec x_\textbf{n})$ contains all the dependence on the inhomogeneous degrees of freedom. Once again, because the real character of the Wheeler-DeWitt equation, the solutions come in conjugated pairs that represent, in terms of the same time variable, a pair of expanding and contracting universes. As we already said, at zero order in the expansion in powers of $\hbar$ of the Wheeler-DeWitt equation, now given by the quantum Hamiltonian constraint (\ref{HC03}), with the semiclassical solutions (\ref{SCWF01}) it is obtained the Hamiltonian constraint (\ref{HJ01}), which in the present case reads
\be\label{HJ02}
-\left( \frac{\partial S}{\partial a} \right)^2 +\frac{1}{a^2} \left( \frac{\partial S}{\partial \varphi} \right)^2 + a^4 V(\varphi) - a^2 = 0 .
\ee
In terms of the time variable $t$ given by (\ref{WKBtime}), which now reads
\be\label{WKBt01}
\frac{\partial}{\partial t} = \pm\left( -\frac{1}{a} \frac{\partial S}{\partial a}\frac{\partial }{\partial a} +\frac{1}{a^3} \frac{\partial S}{\partial \varphi}\frac{\partial }{\partial \varphi} \right) ,
\ee
and implies
\be\label{MOM02}
\dot{a}^2 = \frac{1}{a^2} \left( \frac{\partial S}{\partial a}\right)^2 \ , \ \dot{\varphi}^2 = \frac{1}{a^6} \left( \frac{\partial S}{\partial \varphi} \right)^2 ,
\ee
the Hamiltonian constraint (\ref{HJ02}) turns out to be the Friedmann equation\footnote{Recall that the field $\varphi$ was rescaled according to $\varphi \rightarrow \sqrt{2} \varphi$, see f.n. \ref{fn06}.}
\be\label{FE01}
\left( \frac{\dot a}{a} \right)^2 + \frac{1}{a^2} = \dot \varphi^2 + V(\varphi) .
\ee

At first order in $\hbar$ in the expansion of the Wheeler-DeWitt equation with the semiclassical solutions (\ref{SCWF01}), it is obtained \cite{Kiefer1987, RP2018a} 
\be\label{SCH00}
\mp i \hbar \left( -\frac{1}{a} \frac{\partial S}{\partial a}\frac{\partial }{\partial a} +\frac{1}{a^3} \frac{\partial S}{\partial \varphi}\frac{\partial }{\partial \varphi} \right) \psi = H_m \psi ,
\ee
where the minus sign corresponds to $\phi_+$ in (\ref{SCWF01}) and the positive sign corresponds to $\phi_-$ in (\ref{SCWF01}). The term in brackets in (\ref{SCH00}) is  the time variable of the background spacetime (\ref{WKBt01}), so (\ref{SCH00}) is turns out to be the Schr\"{o}dinger equation for the matter fields that propagate in the classical background spacetime. We have then recovered, at zero and first orders in $\hbar$, the semiclassical picture of quantum matter fields, which satisfy the Schr\"{o}dinger equation (\ref{SCH00}), propagating in a classical spacetime that satisfies the Friedmann equation (\ref{FE01}). However, in order to obtain the correct sign in the Schr\"{o}dinger equation (\ref{SCH00}) one must choose a different sign for the time variables in the two universes of the conjugated pair in  (\ref{SCWF01}). For the branch represented by $\phi_+$ one must take the negative sign in (\ref{WKBt01}) and the positive sign for $\phi_-$. It means that the physical time variables of the two universes, i.e. the time variable measured by actual clocks that are eventually made of matter, are reversely related, $t_2 = - t_1$. Both universes are therefore expanding universes in terms of their physical time variables, $t_1$ and $t_2$ \cite{RP2018c}. Particles propagating in the symmetric universe look as they were propagating backward in time so they can naturally be identified with antiparticles. Thus, primordial matter and antimatter would be created in different universe and that might explain the primordial matter-antimatter asymmetry observed in the context of a single universe \cite{RP2017e}.

\subsection{Semiclassical particles: geodesics and uncertainties in the position}

The analogy between the evolution of the universe in the minisuperspace and the trajectory of a particle in a curved spacetime can make us to ask if the classical trajectories of test particles in general relativity can also be derived from the quantum state of a field that propagate in the spacetime. The answer is yes \cite{Garay2018}. We shall see now that the solutions of the Klein-Gordon equation contain not only information about the matter fields  they represent but also about the geometrical structure of the spacetime where they propagate through the geometrical information contained in the corresponding geodesics. In order to show it, let us consider the analogue in the spacetime to the semiclassical wave function (\ref{WKB01}),
\be\label{SCWFst2}
\varphi(x) = C(x) e^{\pm \frac{i}{\hbar} S(x)} ,
\ee
where, $x = (t, \vec x)$, and $C(x)$ and $S(x)$ are two functions that depend on the spacetime coordinates. Then, inserting the semiclassical wave function (\ref{SCWFst2}) into the Klein-Gordon equation (\ref{KG01}) and expanding it in powers of $\hbar$, it is obtained at zero order in $\hbar$ the following Hamilton-Jacobi equation
\be\label{MCst02}
g^{\mu\nu} \frac{\partial S}{\partial x^\mu} \frac{\partial S}{\partial x^\nu} + m^2 = 0 ,
\ee
which is the momentum constraint (\ref{HCst}) if we make the identification, $p_\mu = \frac{\partial S}{\partial x^\mu}$. Furthermore, with the following choice of the affine parameter,
\be\label{WKBtau}
\frac{\partial }{\partial \tau} = \pm \frac{1}{m} g^{\mu\nu} \frac{\partial S}{\partial x^\mu}\frac{\partial }{\partial x^\nu} ,
\ee 
one arrives at
\be\label{MOMst02}
p_\mu = \pm m g_{\mu\nu} \frac{d x^\nu}{d\tau} .
\ee
With the momentum constraint (\ref{MCst02}) and the value of the momenta (\ref{MOMst02}) one can derive the equation of the geodesic (\ref{GEOst}) (see, for instance, Ref. \cite{Garay2018}). The two possible signs in the definition of the affine parameter in (\ref{WKBtau}) correspond to the two possible ways in which the geodesic can be run, forward and backward in time. These are the solutions used by Feynman to interpret the trajectories of particles and antiparticles of the Dirac's theory \cite{Feynman1949}.

As an example, let us consider the case of a flat DeSitter spacetime, for which the analytical  solutions of the Klein-Gordon are well known. In conformal time, $\eta =\int \frac{dt}{a}$, and in terms of the rescaled field, $\chi(\eta, \vec x)= a(\eta) \varphi(\eta, \vec x)$, the Klein-Gordon equation (\ref{KG01}) becomes
\be
\hbar^2 \chi'' - \hbar^2 \nabla^2 \chi + \left( m^2 a^2 - \hbar^2 \frac{a''}{a} \right) \chi = 0 ,
\ee
where the prime denotes the derivative with respect to the conformal time. Notice here the appearance of the Planck constant with respect of the customary expression of the Klein-Gordon (see, for instance, Refs. \cite{Birrell1982, Mukhanov2007}). Let us go on by decomposing the function $\chi$ in normal modes as
\be
\chi(\eta, \vec x) = \int \frac{d^3\vec k}{(2 \pi)^\frac{3}{2}} \chi_{\vec k }(\eta) e^{\pm \frac{i}{\hbar} \vec k \cdot \vec x} ,
\ee
where the normal modes $\chi_{\vec k}$ satisfy
\be\label{WE01}
\hbar^2 \chi''_{\vec k} + \omega^2_k(\eta) \chi_{\vec k} = 0 ,
\ee
with $k=|\vec k|$, and in the case of a flat DeSitter spacetime
\be
\omega_k^2(\eta) = k^2 + \left( \frac{m^2}{H^2} - 2 \hbar^2 \right) \frac{1}{\eta^2} .
\ee
The solutions of the wave equation (\ref{WE01}) can easily be found \cite{Birrell1982, Mukhanov2007} in terms of Bessel functions. The solution with the appropriate boundary condition is given by \cite{Mukhanov2007}
\be
v_{ k}(\eta) = \sqrt{\frac{\pi |\eta|}{2}} \mathcal H^{(2)}_n\left( \frac{k |\eta|}{\hbar} \right) ,
\ee
where $\mathcal H_n^{(2)}(x)$ is the Hankel function of second kind and order $n$, with
\be
n = \sqrt{\frac{9}{4}-\frac{m^2}{\hbar^2 H^2}} . 
\ee
These are the customary modes of the Bunch-Davies vacuum. Note however the presence here of the Planck constant $\hbar$ in the argument and in the order of the Hankel function. It does not appear when the Klein-Gordon is derived from the action of a classical field. In the present case, it is going to allow us to make an expansion of the modes in powers of $\hbar$. Using the Debye asymptotic expansions for Hankel functions \cite{Garay2018}, one can write
\be
\mathcal H^{(2)}_{\frac{im}{\hbar H}} \left( \frac{k}{\hbar H a} \right) \approx \sqrt{\frac{2\hbar H}{\pi \omega_c}} e^{-\frac{\pi m}{2\hbar H}}  e^{-\frac{i}{\hbar}\left( \frac{\omega_c}{H} - \frac{m}{H} \log\left( \frac{a}{k} (m+\omega_c) \right)  \right) }   \left( 1 + \mathcal O(\hbar) \right) ,
\ee 
where, 
\be\label{DR01}
\omega_c \equiv \omega_{c,k}(\eta)  = \sqrt{ k^2 + m^2 a^2 } .
\ee
Then, the solutions of the Klein-Gordon equation can be written in the semiclassical form of the wave function (\ref{SCWFst}) with,
\be\label{SDS}
S(\eta, \vec x) = \vec k\cdot\vec x -  \frac{\omega_c}{H} - \frac{m}{H} \log\left( \frac{a}{k} (m+\omega_c) \right) .
\ee
In that case, the momentum constraint (\ref{MCst02}) is satisfied because, from (\ref{SDS}), we have
\be
\frac{\partial S}{\partial \eta} = \omega_c(\eta)  ,  \,\,\,\, {\rm and} \,\,\,\, \vec \nabla S = \vec k ,
\ee
so the momentum constraint turns out to be the dispersion relation given by (\ref{DR01}). We can now choose the affine parameter $\tau$, defined by
\be
\frac{\partial }{\partial \tau} = \pm \frac{1}{a^2 m} \left( - \omega_c \frac{\partial}{\partial \eta} + \vec k \cdot \vec \nabla \right)  ,
\ee
in terms of which,
\be
\frac{d \vec x}{d\tau} = \pm \frac{1}{a^2 m} \vec k \,\,\,\, , \,\,\,\, \frac{d \eta}{d\tau} = \mp \frac{1}{a^2 m} \omega_c ,
\ee
that satisfy the geodesic equation of the flat DeSitter spacetime, given by the Euler-Lagrange equations associated to the action (\ref{ACTst}). Therefore, at the classical level, which is given by the limit\footnote{$\hbar$ is a constant so by the limit $\hbar \rightarrow 0$ we mean that the magnitudes at hand are very large when they are compared with the value of the Planck constant.} $\hbar \rightarrow 0$, the solutions of the Klein-Gordon equation give rise to the classical geodesics of the spacetime where they are propagating. It means that the Klein-Gordon equation contains not only information about the quantum state of the field but also about the geometrical structure of the underlying spacetime.

At first order in $\hbar$ it also contains the quantum information given by the Schr\"{o}dinger equation, in the non-relativistic limit. For instance, let us consider normal coordinates ($N=1$ and $N^i, \forall i$ in (\ref{STfol})) so that the metric of the spacetime  becomes
\be
ds^2 = - dt^2 + h_{ij} dx^i dx^j .
\ee
In that case, the Klein-Gordon equation (\ref{KG01}) turns out to be
\be
\hbar^2 \ddot \varphi + \hbar^2 \frac{\dot h}{2 h} \dot \varphi - \hbar^2 \nabla^2_\Sigma \varphi + \left( m^2 + 2m V(\vec x)\right) \varphi = 0 ,
\ee
where we have also consider an external potential, $V(\vec x)$. In the non-relativistic regime, we can assume that the field $\varphi(t,\vec x)$ has the semiclassical form
\be\label{SCWFst}
\varphi(t,\vec x) = \frac{1}{h^\frac{1}{4}} e^{-\frac{i}{\hbar} m t} \psi(t,\vec x) ,
\ee
where $\psi(t,\vec x)$ is the non-relativistic wave function of the field. Then, insert it in the Klein-Gordon equation (\ref{KG01}), and disregarding second order time derivatives, or equivalently orders of $\hbar^2$ and higher, it is obtained \cite{Garay2018} the Schr\"{o}dinger equation for the wave function $\psi(t, \vec x)$, i.e
\be
i \hbar \frac{\partial \psi}{\partial t} = \left( - \frac{\hbar^2}{2 m } \nabla^2_\Sigma + V(\vec x) \right) \psi(t, \vec x) ,
\ee
where $\nabla^2_\Sigma$ is the three-dimensional Laplacian defined in the hypersurface $\Sigma$.

As another example, let us consider now a Schwarzschild spacetime with metric given by
\be
ds^2 = g_{\mu\nu} dx^\mu dx^\nu = - \Delta dt^2 + \Delta^{-1} dr^2 + d\Omega_3^2 ,
\ee
with, $\Delta = 1 -\frac{2M}{r}$, in units for which $c=G=1$. It is now convenient to make the conformal transformation, $\tilde g_{\mu\nu} = \Delta^{-1} g_{\mu\nu}$, 
and the following reparametrisation, $d\lambda = m\Delta^{-1} d\tau$, so that the classical Hamiltonian constraint (\ref{HCst}) can be split into a relativistic part and a non-relativistic part \cite{RP2019a}
\be
H = H_r + H_{nr} = 0 ,
\ee
with 
\be
H_r = -\frac{1}{2 m} p_t^2  +\frac{m}{2}  \,\,\,\, , \,\,\,\,  H_{nr}= \frac{1}{2 m } \tilde h^{ij} p_i p_j + m V(\vec x) , 
\ee
where now, $V(\vec x)=-\frac{M}{r}$, is the Newtonian potencial of a gravitational body with mass $M$, and $\tilde h^{ij}$ is the inverse of the metric induced by  $\tilde g_{\mu\nu}$ in the spatial sections, with $\tilde h_{ij}= \Delta^{-1} h_{ij}$. Far from the Schwarzschild radius, $\Delta \approx 1$, so the metric of the spatial sections induced by $\tilde g_{\mu\nu}$ can be approximated by the metric of the flat space. However, closed to the event horizon  $\tilde h_{ij}$ would entail a significant departure from the flat space. It means that far enough from the gravitational body it is recovered the Newtonian picture of a test particle propagating in a flat spacetime under the action of the gravitational potential $V(r)$.

Quantum mechanically, assuming the value of the semiclassical wave function (\ref{SCWFst}) and following now the procedure explained above, one arrives at the Schr\"{o}dinger equation for the wave function $\psi(t,\vec x)$ with the Newtonian central potential
\be
i \hbar \frac{\partial \psi}{\partial t} = \left( - \frac{\hbar^2}{2 m } \nabla^2_{\tilde\Sigma} + V(r) \right) \psi(t, \vec x) .
\ee 
Therefore, the solution of the Klein-Gordon equation (\ref{KG01}) contains at zero order in $\hbar$, i.e. at the classical level, the classical trajectories of test particles moving in the spacetime where the quantum field propagates and, at first order in the Planck constant, it provides us with the Schr\"{o}dinger equation that gives the dispersion in the position of the test particle with respect to the classical trajectory trough the well-know relation
\be
\Delta \vec x = \langle \psi | \hat{\vec x}^2 | \psi \rangle - \langle \psi | \hat{\vec x} | \psi \rangle^2 .
\ee
Therefore, the modes $u_k(x)$  of the second quantisation procedure given in Sec. \ref{sec04} represent, in the semiclassical regime, particles that propagate with high probability along the geodesics of the spacetime but with a given uncertainty in their positions given by the Schr\"{o}dinger equation. Of course, the particle interpretation of the modes is only valid for modes for which the wavelength is significantly less that the characteristic length of the particle detector. However, it provides us with a clear picture for the interpretation of the quantum field. 

Similarly, the modes $u_i(q)$ of the third quantisation procedure, which are the solutions of the Wheeler-DeWitt equation, represent semiclassical universes in the sense that they represent, at zero order in $\hbar$, the classical spacetime background where the matter fields propagate and, at first order in $\hbar$, the uncertainties in the values of the matter fields. Therefore, the wave function $\phi(a,\varphi)$, which can be seen as a field that propagates in the minisuperspace, represent the quantum state of a field that is made up of universes with matter contents that are randomly distributed among all the possible values. It represents therefore the quantum state of the whole multiverse, in the minisuperspace approximation.


\section{Conclusions and further comments}\label{sec06}

There is a formal analogy between the evolution of the universe in the minisuperspace and the trajectory of a test particle in a curved spacetime that allows us to interpret the former as a trajectory in the minisuperspace with parametric coordinates given by the solutions of the classical field equations, $a(t)$ and $\varphi(t)$. The time variable $t$ is  the parameter that parametrises the trajectories. The invariance of the Lagrangian associated to the Hilbert-Einstein action, and therefore of the field equations too, with respect to a time reversal change of the time variable indicates that the universes must be created in pairs with opposite values of the momenta conjugated to the minisuperspace variables. Thus, the creation of the universes would also conserved the total momentum. A positive value of the momentum conjugate to the scale factor  entails a positive value of the zero component of the tangent vector to the trajectory, i.e. it  entails an increasing value of the scale factor so it represents an expanding universe. In terms of the same time parametrisation, the partner universe with the opposite value of the momentum  entails a decreasing value of the scale factor so it represents a contracting universe. Therefore, in terms of the same time variable the universes are created in pairs, one contracting  and the other expanding.

The analogy between the evolution of the universe in the minisuperspace and the trajectory of a test particle in the spacetime can be extended to the quantum picture too. The wave function that represents the quantum state of the spacetime and the matter fields, all together, can be seen as a super-field that propagates in the minisuperspace. Then, a third quantisation procedure can be applied that parallels that of the second quantisation of a field that propagates in the spacetime. We can then define creation and annihilation operators of universes as well as a Fock space for the state of the super-field, which can be then interpreted as made up of universes evolving (i.e. propagating) in the minisuperspace. The appropriate representation to describe the universes in the minisuperspace is the invariant representation of the quantum Hamiltonian associated to the Hilbert-Einstein action. In terms of the invariant representation the ground state of the super-field represents the \emph{nothing} state, which corresponds to the state of no universe at all at any point of the minisuperspace. However, the minisuperspace could be full of universes if the boundary state of the super-field is  the ground state of a different representation. In particular, if the boundary state of the super-field is the ground state of the diagonal representation of the Hamiltonian at some value $a_0$, which is the value of the scale factor at which the universes are created, then, the minisuperspace would be full of pairs of universes with opposite values of their momenta conjugated to the variables of the minisuperspace in a perfectly correlated or entangled state.

In the semiclassical regime we recover the picture of quantum matter fields propagating in a classical spacetime background. The modes of the mode decomposition of the super-field represent, in that case, semiclassical universes propagating in the minisuperspace. The cosmic time naturally appears in this regime as the WKB parameter that parametrises the classical trajectory, i.e. it parametrises the classical evolution of the spacetime background of the universes. At first order in the Planck constant, we obtain the Schr\"{o}dinger equation that determines the quantum evolution of the matter fields in a pair of universes. However, the time variable in the two universes of the pair must be reversely related in order to obtain the appropriate value of the Schr\"{o}dinger equation in the two symmetric universes. It means that in terms of their physical time variables, i.e. in terms of the time variables given by actual clocks that are eventually made of matter, the two universes of the symmetric pair are both expanding or contracting. The consistent solution would be considering two expanding universes because two newborn contracting universes would rapidly delve again into the gravitational vacuum from which they just emerged. For an internal inhabitant of the universe, the particles that move in the partner universe would look like if they were propagating backward in time so they would naturally be identified with the antiparticles that he or she does not see in his/her universe. The matter-antimatter asymmetry observed in the context of a single universe would be thus restored.

The semiclassical formalism can be applied to the quantum state of a field that propagates in a curved spacetime. In that case, the zero order component in $\hbar$ of the semiclassical expansion of the field gives rives to the classical equation of the geodesic of the underlying spacetime. Therefore, the solutions of the Klein-Gordon equation contain not only information about the quantum state of the field but also information about the geometrical structure of the spacetime where it propagates. At first order in $\hbar$ one obtains the corresponding Schr\"{o}dinger equation that drives the uncertainties in the position of the particles of the field. Therefore, the field is well represented in the semiclassical regime by classical particles propagating with the highest probability along the geodesics of the spacetime but with some uncertainty or deviation from the classical path given by the wave functions of the corresponding Schr\"{o}dinger equation.

\section*{Acknowledgments}

I would like to thank the organisers of the meeting \emph{Travelling through Pedro's universes}, held in Madrid at the Universidad Complutense the days $3^{rd}-5^{th}$ of Dec., for the memory and the tribute made to the figure of Pedro F. Gonz\'{a}lez-D\'{\i}az, who was my thesis supervisor and the scientific reference of all my work, as well as a very good friend; and for inviting me to give the talk on which this paper is based.


\bibliography{bibliography}

\end{document}